\documentclass[twocolumn,showpacs,aps,prl]{revtex4}
\usepackage{epsfig,amssymb}

\begin{document}

\title{Decoherence-driven Cooling of a Degenerate Spinor Bose Gas}
\author{H.~J. Lewandowski$^\ast$, J.~M. McGuirk$^\ast$, D.~M. Harber, and E.~A. Cornell$^\ast$}
\affiliation{JILA, National Institute of Standards and Technology and\\ University of Colorado and Department of Physics, 
University of Colorado,\\ Boulder, Colorado 80309-0440}
\date{\today}

\begin{abstract}
We investigate the relationship between the coherence of a partially Bose-condensed spinor gas and its temperature. We observe cooling of the normal component driven by decoherence as well the effect of temperature on decoherence rates.  
\end{abstract}

\maketitle

Understanding the effects created by the loss of coherence in Bose-condensed systems is critical if condensates are to be used in applications such as quantum information or precision measurement. In quantum information systems, the loss of coherence limits the duration of an experiment, and thus the complexity of the computation. The loss of coherence will also limit the precision of spectroscopic measurements of small energy shifts (\textit{e.g.}, caused by Casimir and hypothetical short-range forces \cite{dimopoulos}).

The extent to which finite temperature erodes the deterministic nature of the phase of a condensate over spatial displacements has been studied in Ref. \cite{ertmer2001}.  Measuring temporal coherence is typically technically more challenging, and, with the exception of Ref. \cite{hall1998}, measured coherence times in condensates are typically less than 5 ms \cite{shin}. In order to observe condensate coherence times in excess of 100 ms, the JILA group \cite{hall1998} made use of the convenient properties of a spinor gas system, with a nearly pure condensate. In this Letter we study the relationship between the temperature of a spinor gas and the evolution of its coherence over time. As we shall see, the arrow of causality points both ways, and we discuss, in order, experiments showing that (i) decoherence can lead to temperature change and (ii) differing temperatures strongly affect decoherence rates.

The first part of this Letter deals with spontaneous cooling, driven by decoherence, of the normal component in a partially condensed system.  The phenomenon of decoherence-driven cooling can be described as follows: Consider a partially condensed system in which the condensate and normal component are both in the same fully coherent superposition of two spin states. In effect, this coherent superposition represents a single species of indistinguishable atoms. Now if the normal component decoheres, due to an inhomogeneous potential for instance, the atoms in the normal component become distinguishable particles and must be thought of as atoms in two distinct populations. There will now only be half as many atoms in the normal component as are required by Bose statistics to support a condensate of the initial size. The decoherence essentially reduces the effective quantum-state occupation number by two. The condensate must then transfer atoms to the normal component to restore thermodynamic equilibrium. 

The atom flux out of the condensate is not a result of increasing the energy in the system through heating, but rather a consequence of the changing statistics. In fact conservation of energy implies that the atom flux is accompanied by a subsequent cooling of the normal component. The condensate transfers atoms to the normal component with nearly zero energy. The initial thermal energy is then redistributed among the now larger number of normal atoms, lowering the temperature of the normal component, so as to restore the normal-component phase-space density to its saturated value of 2.61. 

A measurement of decoherence-driven cooling of the system requires that the normal component must decohere on a time scale shorter than the lifetime of the condensate. The most obvious states for measuring coherence in $^{87}$Rb are the $|1\rangle \equiv |F=1,m_f=-1\rangle$ and $|2\rangle \equiv |F=2,m_f=1\rangle$ of the  5S$_{\scriptsize{\mbox{1/2}}}$ manifold. However, we find that the coherence time of the local spin \cite{localspin} of the normal component in a superposition of these states ($\sim$100 ms) is not significantly shorter than the lifetime of the condensate ($\sim$150 ms) for a cloud with 35$\%$ of the atoms in the condensate. Decoherence of the normal-component spin is driven by inhomogeneity in mean-field and Zeeman shifts, and by interactions with the condensate. The differential Zeeman potential experienced by the $|1\rangle$ and $|2\rangle$ states can be modified by changing the magnetic bias field of the trap \cite{harber2002}, but it is difficult to reduce the coherence time of the normal component significantly below the lifetime of the condensate. In particular the presence of the effects of spin waves, which stiffen the spin field, delays the onset of decoherence \cite{harber2002,mcguirk2002}.

To satisfy the condition on the coherence time we instead use dressed states, which are coherent equal superpositions of the $|1\rangle$ and $|2\rangle$ states in the presence of a resonant dressing field \cite{cohen1996}. The dressed states are eigenstates of the system when the dressing drive is applied. As in the bare-state system, in the dressed-state system the differential potential determines the coherence time for the normal component. The differential potential in the dressed-state system is determined by the spatial Rabi frequency inhomogeneity produced by a gradient in the dressing field strength. We are able to produce a large enough inhomogeneity in the Rabi frequency, on the order of 100 Hz across the cloud, to cause the normal component to decohere in $\sim$4 ms, much less that the condensate lifetime of $\sim$150 ms. Another feature of the dressed-state system is the near perfect symmetry between the two states. Resonantly dressed states are equal superpositions of the bare states, giving identical collisional and loss properties, and thus equivalent condensate lifetimes. The details of the dressed-state system are not critical for the comprehension of the these experiments. Analogous to the bare-state system, the dressed states just form a two-level quantum system. 

The experimental apparatus is described in detail in Refs. \cite{lewando2002,lewando2003} and will be summarized here. $^{87}$Rb atoms are pre-cooled in a magneto-optical trap and transferred to an Ioffe-Pritchard style magnetic trap via a moving quadrupole magnetic trap. The axial and radial Ioffe-Pritchard trapping frequency are 7 Hz and 230 Hz respectively. Atoms in the $|1\rangle$ state are cooled by radio frequency evaporation to below the Bose-Einstein condensation transition temperature. From here the atoms are placed in either a single dressed state or a fully coherent superposition of dressed states by controlled application of a two-photon microwave dressing field \cite{dressing}.

\begin{figure}
\leavevmode
\epsfxsize=3.375in
\epsffile{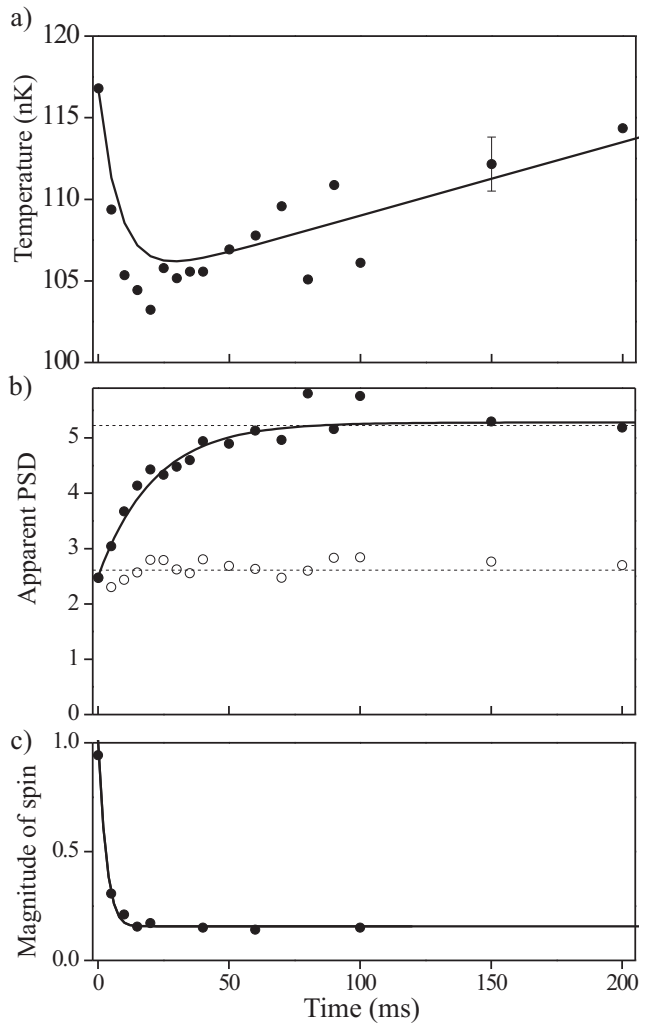}    
\caption{\label{hjlfig1}a) Temperature of the normal component, which is initially in a superposition of states, as a function of time. The temperature initially decreases as the condensate contributes atoms with nearly zero energy to the normal component. At later times the temperature increases due to heating from inelastic collisions. The solid line is a model as explained in the text. The single error bar is representative of the statistical error on all points. b) The apparent phase-space density (PSD) is shown as a function of time. The open circles are the PSD of the normal component in a single state, whereas the solid circles represent the PSD of the normal component in an initially fully coherent superposition of states. The superposition-state PSD approaches 2 $\times$ 2.61 in a time scale of 21(3)~ms.  Each point is a weighted average of six independent measurements. The standard deviations of the six measurements are equal to or smaller than the point size. The dashed lines are at PSDs of 2.61 and 5.22. c) Magnitude of the spin vector as a function of time for the normal component initially in a fully coherent superposition of states. The entire spin vector is reconstructed from measurements of the longitudinal and transverse spin components similar to those described in Ref. \cite{mcguirk2003}. The coherence time of the normal component is 3.9(4) ms. The offset in the value of the spin vector from zero at long times is due to imaging and shot-to-shot number noise.}
\end{figure}

Initially we create a partially condensed sample with 35$\%$ of the atoms in the condensate, with both the normal component and condensate in a fully coherent equal superposition of dressed states. We allow the system to evolve for some time, during which the normal component rapidly decoheres, and then image the sample using absorptive imaging.  
The image is fitted using a Thomas-Fermi profile for the condensate and a Gaussian profile modified by Bose statistics for the normal component from which the temperature is extracted \cite{Ketterle1999a}. The resulting temperature is plotted as a function of time in Fig. \ref{hjlfig1}a.

As seen in Fig.~\ref{hjlfig1}a the temperature in the normal component rapidly decreases as condensate atoms are transferred to the normal component to maintain thermodynamic equilibrium. Around 30 ms the condensate atom ``melting'' tapers off as the system approaches equilibrium, and the temperature rises due to heating from inelastic collisions. The solid curve in Fig. \ref{hjlfig1}a is a simple model for the normal component temperature. The inputs to the model include an empirically determined number of condensate atoms transferred to the normal component and a measured heating rate. The heating rate as well as a loss rate from two-body spin relaxation is determined by initially placing the cloud in a single dressed state and measuring the temperature and number as a function of time. The energy in the normal component is thermally redistributed among all of the atoms in the normal component, and a new temperature is calculated. 

Phase-space density (PSD) is another parameter that elucidates the loss of correlations from decoherence. A normal component in a single state or a coherent superposition of states (\textit{i.e.} a single quantum state) in thermal equilibrium with a condensate has a PSD of 2.61 as specified by Bose statistics. Correspondingly two distinguishable species in a single trap, $^{87}$Rb and $^{85}$Rb for instance, each have a normal component with a PSD of 2.61.  This can be experimentally verified by measuring the number and temperature \textit{independently} of the $^{87}$Rb and $^{85}$Rb isotopes. However if one counted the total number of atoms and measured the size of the cloud regardless of isotope, the calculated apparent PSD would be 2 $\times$ 2.61. The unphysical PSD of 5.22 is due to incorrect counting of quantum states in the system.

Suppose that instead of two atomic species one has two spin states in $^{87}$Rb. A sample in a fully coherent superposition of the two states will always give a PSD of 2.61. On the other hand an incoherent mixture of these two states can give an apparent PSD of 5.22 if the measurements are insensitive to the internal atomic state. For this reason we can use apparent PSD as a measure of correlations (coherence) of the normal component.

We calculate the apparent PSD for a cloud in an initially fully coherent superposition of states from the temperature and number extracted from the same images taken for the cooling measurement. To remove fitting systematics, we also measure the temporal evolution of the phase-space density of a cloud in a single dressed-state. There is a systematic error in calculating the phase space density of the normal component when a condensate is present due to our fitting routine, which does not take into account the reduction of the normal component peak density from the mean-field pressure of the condensate. We remove this systematic error by fitting the single-state phase-space density versus condensate fraction data and removing the same fitted trend from both the single- and superposition-state data. 

We are able to observe the apparent phase-space density of the superposition case evolve from 2.61$\rightarrow $5.22 (Fig.~\ref{hjlfig1}b).  The time scale for the apparent PSD to increase is $\sim$ 20 ms, as extracted from an exponential fit to the data in Fig. \ref{hjlfig1}b. The normal component becomes incoherent on a much faster time scale ($\sim$ 4 ms) (Fig. \ref{hjlfig1}c) than the system can equilibrate, which is limited by the elastic collision rate. The radially averaged elastic collision rate for the data in Fig. \ref{hjlfig1} is $\sim$ 200 Hz and several collisions are required to enforce equilibrium.

In the second part of this Letter we now proceed to discuss a series of measurements we make characterizing the
temperature dependence of spin decoherence rates in partially condensed systems, working now in our original system of bare spin states, rather than in dressed states. Mechanisms that affect coherence in purely normal clouds have been studied in Refs. \cite{harber2002} and \cite{mcguirk2002} and include (i) inhomogeneity in the relative potentials for the two spins, an effect which drives decoherence, and (ii) the phenomenon of spin-waves, an effect that arises from spin-exchange collisions, which tends if anything to suppress decoherence.  The addition of a condensate component to the system impacts normal-component coherence in two distinct ways. In the first, more straightforward, effect the condensate is a high-density feature in the middle of the normal cloud that contributes significantly to the spatial inhomogeneity in the mean-field frequency shift. In the second effect, exchange collisions with the condensate tend to induce spin-locking, whereby the spin-fields of the normal cloud and of the condensate are locally aligned \cite{mcguirk2002}.  Normal component-condensate coherence in a finite-temperature spinor gas are discussed with some rigor in Ref. \cite{williams2003}; here we argue qualitatively that spin-locking can lead to either an enhancement or a suppression of normal-cloud coherence, depending on experimental conditions.

In the limit of a relatively small normal cloud, a given normal-component atom spends a large fraction of its time
immersed in the condensate. During this time, its spin's transverse phase is kept well locked to the condensate's. When it emerges from the condensate, it spends relatively little time away before returning to the condensate region. During this time, its transverse phase evolves relatively little, and when it returns to the condensate region, its spin relocks with the condensate with relatively little increase in entropy. In this limit, the condensate acts as a reservoir of coherence, and the spin-locking effect extends the coherence time of the normal cloud.  In the
opposite limit, the normal atom spends significant time away from the condensate, experiencing a differing relative potential and spin-exchanging with its normal brethren. When it returns to the condensate region its spin has evolved to be significantly dephased from the condensate, so that the normal component-condensate spin-locking occurs only at the cost of significant decrease in the magnitude of the normal atom's transverse spin.

\begin{figure} 
\leavevmode
\epsfxsize=3.375in
\epsffile{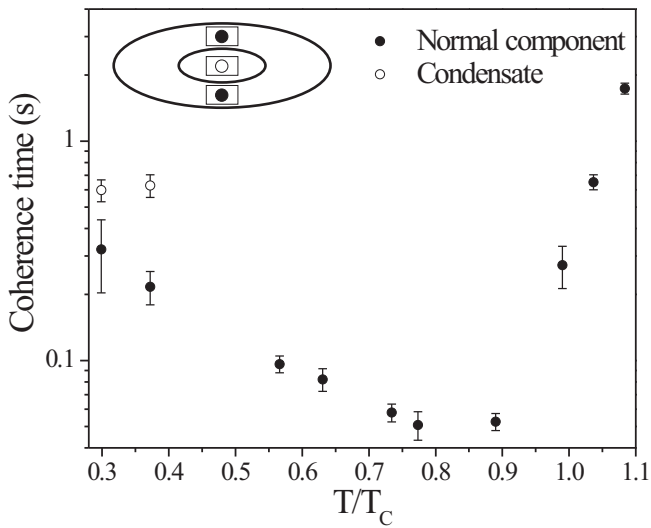} 
\caption{\label{hjlfig2}Coherence times of the condensate region ($\circ$) and the normal component ({$\bullet$}) as a function of T/T$_{\scriptsize\mbox{C}}$. The condensate and normal component regions are defined using the Thomas-Fermi radius, R$_{\scriptsize\mbox{TF}}$. The rectangular regions in the sketch of the expanded density profile (not to scale) indicate the regions over which the local coherence of the normal and of the condensate components are probed. The axial (horizontal) extent of the rectangles is 4 $\mu$m. }
\end{figure}

To measure the coherence times of the normal and condensate components we use standard Ramsey spectroscopy on the $|1\rangle$ and  $|2\rangle$ states \cite{ramsey1956}. We start with the atoms in the $|1\rangle$ state and apply a $\pi$/2 pulse, which creates a coherent equal superposition of the $|1\rangle$ and $|2\rangle$ states. Next we allow the system to evolve and apply a second $\pi$/2 pulse to measure the phase and magnitude of the transverse spin. We then expand the cloud and separately image each spin state. Figure~\ref{hjlfig2} indicates schematically the regions in the cloud over which the optical depth is averaged to determine the local spin coherence in the normal and condensate components. The phase of the coupling drive is swept relative to the first pulse to scan a Ramsey fringe. The degree of coherence, and the time-scale for its decay, is determined by fitting sinusoids to single Ramsey fringes collected at successively longer times. We fit the fringe contrast to a decaying exponential in time. The 1/e decay time is plotted as a function of temperature T, normalized by transition temperature, T$_{\scriptsize\mbox{C}}$, in Fig. \ref{hjlfig2}

It is interesting to note that there is over an order of magnitude reduction in the coherence time from a normal component without a condensate to a cloud with a small condensate present (Fig. \ref{hjlfig2}). The coherence of the normal component rapidly decreases as the temperature is lowered below T$_{\scriptsize\mbox{C}}$. As described above, the loss of normal-cloud coherence is accelerated by a combination of the effects of the increased spatial inhomogeneity associated with the condensate density spike and of non-adiabatic spin exchange with the condensate.

Different behavior is seen when T/T$_{\scriptsize\mbox{C}} \ll 1$. In this case nearly all of the normal atoms are in constant contact with the condensate via exchange collisions.  The coherence time of the normal component is not as drastically decreased but rather has its coherence maintained by the condensate through spin-locking. 

The coherence of the condensate can not be accurately determine when there is a large normal component present. For values of T/T$_{\scriptsize\mbox{C}}$ near unity there are significant numbers of normal atoms present in the region of cloud deemed the condensate region (Fig. \ref{hjlfig2}). The imaging procedure does not distinguish between condensed and non-condensed atoms within the condensate region (0.9$ \times$ R$_{\scriptsize\mbox{TF}}$ ); therefore the measured coherence time does not accurately represent the coherence of the condensate. However for small T/T$_{\scriptsize\mbox{C}}$ we measure that the coherence time of nearly pure condensates is over 0.5 s, which is on the order of the lifetime of the condensate in the $|2\rangle$ state and is the longest reported condensate coherence time.

In conclusion we have observed thermodynamic effects driven by decoherence in a partially condensed spinor system including normal-component cooling and increasing apparent phase-space density. We have also measured the temporal coherence of both the normal component and condensate as function of temperature, giving a condensate coherence times in excess of 0.5 sec. Armed with the knowledge of temporal coherence in partially condensed systems gained from these experiments, one can now use similar spinor systems for precision measurements.

This work was supported by the NSF and NIST.


\begin{thebibliography}{H!}
\bibitem{localspin}
The coherence of the sample is defined locally as an average of the spin in a small region 4 $\mu$m wide in the axial direction in the center of the cloud. If the spin were averaged over the entire cloud the coherence time would be shorter.
\bibitem{dressing}
We can place the atoms in either a single dressed state or a superposition of dressed states by adiabatically or diabatically ramping the frequency of the dressing field onto resonance. Similarly we can project the atoms into either the bare- or dressed-state basis for imaging.
\end{thebibliography}
\end{document}